\documentclass{article}
\usepackage[utf8]{inputenc}

\usepackage{geometry}
\usepackage{graphicx}
\usepackage{enotez}
\let\footnote=\endnote

\title{When negative is not ``less than zero'': Electric charge as a signed quantity}
\author{Alexis Olsho, Suzanne White Brahmia, Andrew Boudreaux, and Trevor Smith}
\date{}

\begin{document}

\maketitle

Electromagnetism (E\&M) is often challenging for students enrolled in introductory college-level physics courses. Compared to mechanics, the mathematics of E\&M is more sophisticated and the representations are more abstract. Furthermore, students may lack productive intuitions they had with force and motion. In this article, we explore the mathematization of electric charge. Specifically, we explore how difficulties with positive and negative signs can arise for learners who approach integers primarily as positions on a number line. In sections \ref{sec:QuantifyingQ} and \ref{sec:SignedQuant}, we discuss the nuances of electric charge as a physical quantity and situate it in a body of work by mathematics and physics education researchers to characterize the uses and meanings of signs. In section \ref{sec:Investigation}, we describe preliminary research that illustrates the effect of wording differences on student reasoning about electric charge as a signed quantity. Finally, in Section \ref{sec:InstrImplications}, we discuss implications for instruction.

\section{Electric charge as a physical quantity}
\label{sec:QuantifyingQ}
Physicists \emph{mathematize} the world---that is, physicists create mathematical quantities and relationships to analyze and explain real-world phenomena \cite{brahmia2019b}. \textit{Quantification} of electric charge (that is, the representation of electric charge as a quantity) is an example of a specific---perhaps idiosyncratic---use of sign in physics, with ``positive'' and ``negative'' acting as \emph{labels} for different charge states of matter. Indeed, as Arons points out, ``the names are perfectly arbitrary and could just as well have been chosen to be `red' and `blue' \ldots or `charming' and `revolting'" \cite{arons1990}. The choice of which type of charge to associate with the electron is arbitrary in a further sense---the quantity $-5~\mu\textrm{C}$ is not inherently negative; instead, the ``$-$'' simply indicates which type of charge is present, while the number (with unit) indicates the amount of that type that is present. 

Consider the two statements ``$-5~^{\circ}\textrm{F} < +5~^{\circ}\textrm{F}$'' and ``$-5~\mu\textrm{C} < +5~\mu\textrm{C}$.''  The former is unambiguous:  temperatures are conceptualized using a number line and its implied ordering of values. The negative sign signifies a value lower than an established reference, while the positive sign signifies a higher value.  The values $-5~^{\circ}\textrm{F}$ and $+5~^{\circ}\textrm{F}$ are not symmetric in a physically meaningful way;  using the Celsius scale, the numerical values would no longer have the same absolute value, and would have the same sign.  The statement ``$-5~^{\circ}\textrm{F} < +5~^{\circ}\textrm{F}$'' simply indicates one temperature is higher than the other.

The statement ``$-5~\mu\textrm{C} < +5~\mu\textrm{C}$,'' however, introduces ambiguity. The negative and positive signs here signify which of two types of electric charge is in surplus.  $-5~\mu\textrm{C}$ and $+5~\mu\textrm{C}$ denote symmetric states, with the same amount of surplus of the different types of electric charge, while the value 0 would correspond to the physical state of balance of equal amounts of the two types.  The two symmetric charge states would still be represented by numbers with the same absolute value if electric charge were operationalized with a different unit of measure.  To a physicist, the best answer to the question ``Which of these two objects has more net charge?'' might be ``Neither!''  We thus argue that  $-5~\mu\textrm{C}$ and $+5~\mu\textrm{C}$ are better conceptualized as equal amounts of departure from balance, rather than values ordered on a number line.  We further argue that the statement ``$-5~\mu\textrm{C} < +5~\mu\textrm{C}$,'' due to its close association with number line ordering, can obscure the physically meaningful insight that $-5~\mu\textrm{C}$ and $+5~\mu\textrm{C}$ represent states with the same amounts of unbalanced charge.

We offer an analogy to color charge.  In quantum chromodynamics, each type of color charge (e.g., red) has a corresponding negative (antired).  A particle with a red color charge of +5 units thus sums to zero net color charge with its anti-particle, of red color charge $-5$ units. We might think of the first particle as ``more red'' than the second, and the second as ``more anti-red'' than the first, but would regard them to have equal amounts of color. For electric charge, we might say that a negative charge is ``less positive'' than a positive charge, but not that it ``has less charge'' than a positive charge. This suggests that for temperature, the sign does not signify any fundamental opposition of complementary types, but rather an ordering along a number line, with $0$ denoting a reference. 

Positive and negative signs are well-suited as labels for electric charge. Because charge is conserved, and the two types of charge are complementary, positive and negative charges behave as real numbers under addition and subtraction. For example, the equation $0 - (-5~\mu\textrm{C}) = +5~\mu\textrm{C}$ concisely describes the removal of $-5~\mu\textrm{C}$ from an electrically neutral object, leaving the object with a net charge of $+5~\mu\textrm{C}$. Moreover, the quantitative statement of Coulomb's Law ($\vec{F}_{1,2}=\frac{kq_1q_2}{r^2}\hat{r}$) exploits the multiplicative properties of positive and negative numbers to to express the empirical rule ``like charges repel, and unlike charges attract.''

Despite the correspondence between electric charge and real numbers, care must be taken when discussing \emph{net charge}. An object with a net charge of $-5~\mu\textrm{C}$ typically contains a relatively large amount of balanced positive and negative charge, and a small amount of unbalanced negative charge. (Note that here we use ``amount'' to refer to an inherently positive quantity; i.e., it is possible to have ``an amount of negative charge'' but not ``a negative amount of charge.'')  On this basis, we offer a working definition of ``net charge'': an object with net charge $+Q$ has an amount $Q$ of unbalanced charge of the positive type, while an object with net charge $-Q$ has an amount $Q$ of unbalanced charge of the negative type. (Q now represents a positive number, rather than a signed number---this definition is making explicit the use of a sign to represent one of two types, and the use of a positive number to represent an amount.) In most physics contexts, the amount of balanced positive and negative charge is unknown and unimportant. A change from electrically neutral to a non-zero net charge \emph{of either sign} corresponds to an \emph{increase} in the amount of unbalanced charge.

\section{Reasoning about signed numbers}
\label{sec:SignedQuant}
 Our investigation of student reasoning about signed quantities in physics (such as electric charge) has been informed by the work of mathematics education researchers. Vlassis studied the development of ``flexibility'' with the negative sign, finding that understanding and applying different meanings of the negative sign was correlated with ability to solve linear equations with one unknown \cite{vlassis2004}. Vlassis synthesized these meanings in a map of the ``natures of negativity'' in algebra. Inspired by Vlassis's map, and motivated by the relative lack of research on negative \textit{physics} quantities, we undertook development of a framework for the natures of negativity in physics \cite{brahmia2020}. 
In creating the framework, we identified the use of sign as an \textit{identifier of type} as unique to physics. 
 
Bishop \textit{et al.} identified productive strategies in students' reasoning about negative numbers prior to formal instruction \cite{bishop2014}. Two of these strategies interested us in particular: number-line reasoning, which ``leverages the sequential and ordered nature of numbers,'' and magnitude reasoning, which relates numbers (including negative numbers) ``to a countable amount or quantity'' \cite{bishop2014}. As discussed above, ``net charge'' can be understood as the amount of unbalanced charge present, where the sign specifies the type of charge in surplus. This way of thinking is well-served by the magnitude-based reasoning strategy for negative numbers, which is associated with ``the view of a number having magnitude or substance.'' In this approach, negative numbers may ``evoke the idea of opposite (directed) magnitudes'' \cite{bishop2014}. This contrasts with the number-line-based reasoning strategy, in which students consider quantities as ordered positions on a number line. 

When using a number line to support addition and subtraction, ``one typically treats the start and result as locations on the number line and the change as a distance'' \cite{bishop2014}.  We do not assert that using a number line to aid in adding or subtracting would result in incorrect calculations about electric charge. However, ordering reasoning may lead students to treat a net charge of -5 units as intrinsically less than a net charge of 0 units.  The student might then plausibly fail to consider the implicit meaning of the sign as a signal for which of the two types of charge is in surplus. While further research could reveal whether or not such confusion is indeed prevalent, we here simply identify the possibility that the ubiquitous ordering reasoning associated with positive and negative values arrayed on a number line could ``crowd out'' the desired physical reasoning involving two distinct types of electric charge.

\section{Investigation of student reasoning about electric charge as a signed quantity}
\label{sec:Investigation}

Our investigation of student reasoning about electric charge as a signed quantity began with the development of a multiple-response test item, the Charged Spheres question, as part of a suite of items designed to investigate student interpretation of negative signs in physics contexts \cite{brahmia2017a}. 
The Charged Spheres question proved challenging for introductory students, and also, during expert validation of the item, for physics graduate students.

\begin{figure}
\begin{center}
\fbox{%
\includegraphics[width=0.95\textwidth]{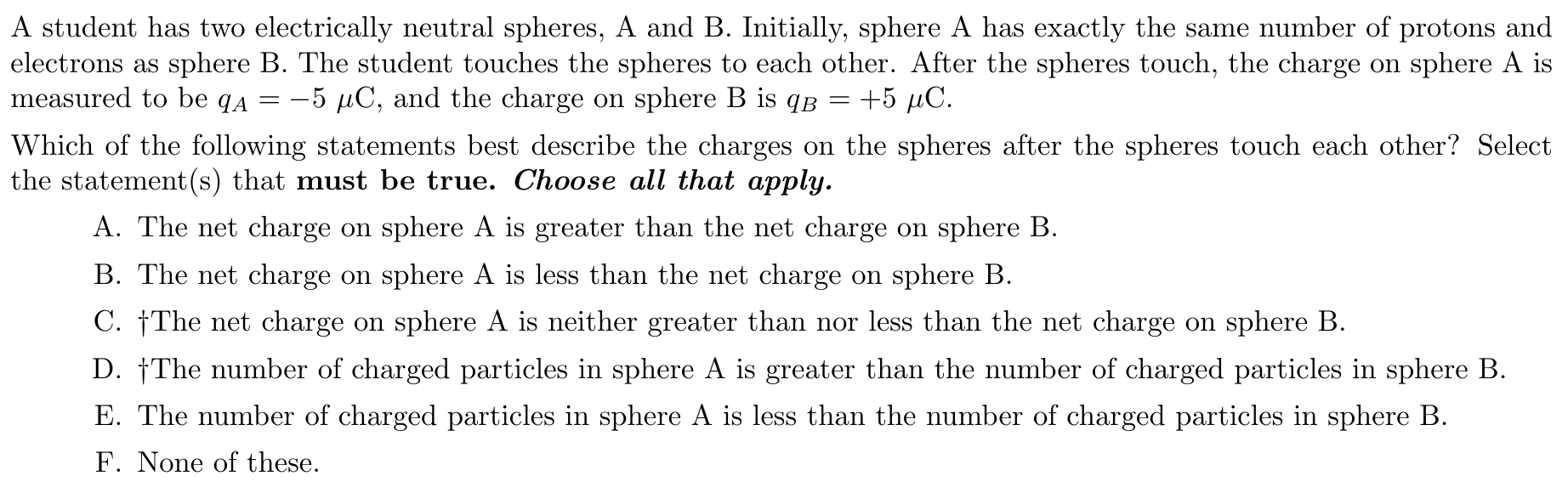}
}
\fbox{%
\includegraphics[width=0.95\textwidth]{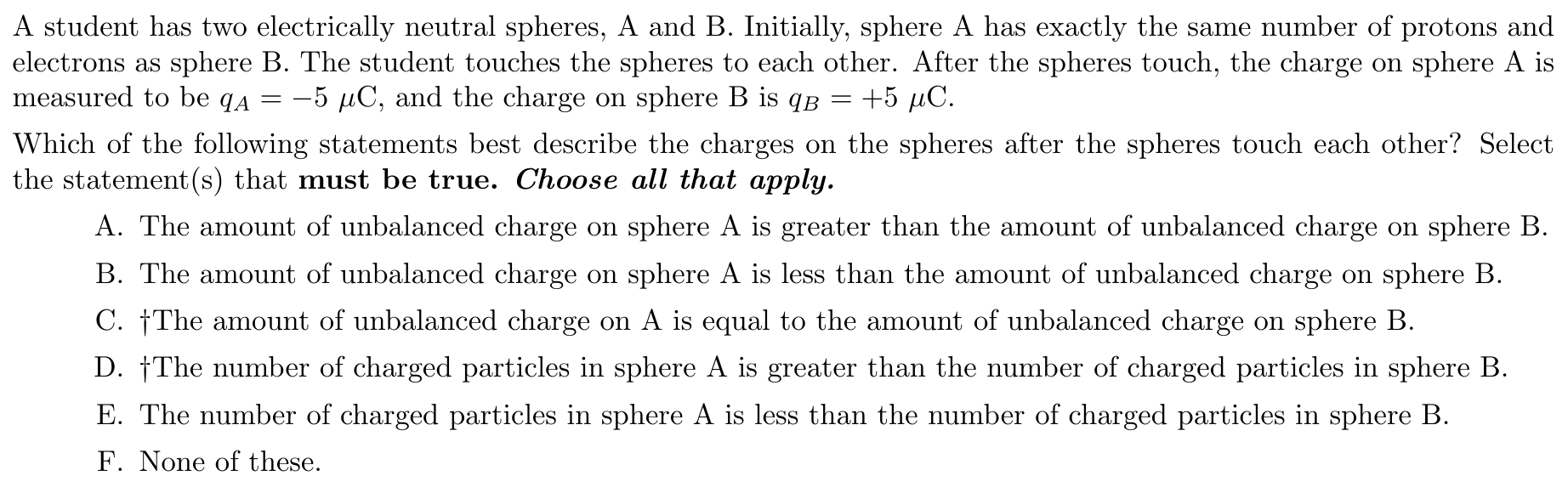}
}

\caption{Versions 1 (top) and 2 (bottom) of the \textit{Charged Spheres} questions, intended to probe student reasoning about net charge and ``negativity'' in charge. Responses C and D are both correct for both versions (shown with daggers).}
\label{fig:ChargedSpheres}
\end{center}
\end{figure}

To investigate further, we modified the item, creating the two different versions shown in Figure \ref{fig:ChargedSpheres}.   These two versions involve the same physical context and wording of the question stem, but use different wording for the answer choices. In the first version, the answer choices refer to ``net charge,'' while in the second version, the answer choices refer to ``the amount of unbalanced charge.'' For both versions, answer choice B is consistent with number-line reasoning---which treats  a negative number as  less  than zero---while  choice C is  consistent  with  magnitude-based reasoning---which  relates  a negative number to a countable amount of something. On both versions, selecting both C and D is the correct response.  

We used the modified Charged Spheres question to explore which wording, if either, might be associated more strongly with the interpretation of sign as an indicator of type in the context of electric charge.  We wondered whether the more descriptive language of the second version would more strongly cue a comparison of two positive quantities (i.e., the amount of surplus charge on Sphere A and the amount on B), and, correspondingly, whether this more descriptive language might suppress reasoning associated with order on a number line. We emphasize that answer choice C in the first version of the question (``the net charge on Sphere A is neither greater than nor less than the net charge on Sphere B'') does not imply that the net charge of the spheres are equal. Validation interviews suggested that students were interpreting the question as intended. Students choosing C typically explained that ``$-$'' and ``$+$'' are used as labels for charge type.  Moreover, students choosing B justified this answer using reasoning suggesting an overarching belief that ``negative is less than positive'' for net charge.
 
The Charged Spheres question was administered as part of an online, ungraded pretest in the second quarter of the calculus-based introductory physics sequence at a large public research university. The students were enrolled in three different sections of the same course, with question version being randomly assigned by section. There were no significant differences in the sections’ average midterm or final exam scores. Though the sections had different instructors, instruction was standardized, and all lecturers used the term ``net charge'' or simply ``charge'' during lecture, consistent with the course textbook. The task was administered after all relevant lecture instruction on charge.  Each student saw only one version of the question.
 
\begin{figure*}
    \centering
    \includegraphics[width = 0.45\textwidth]{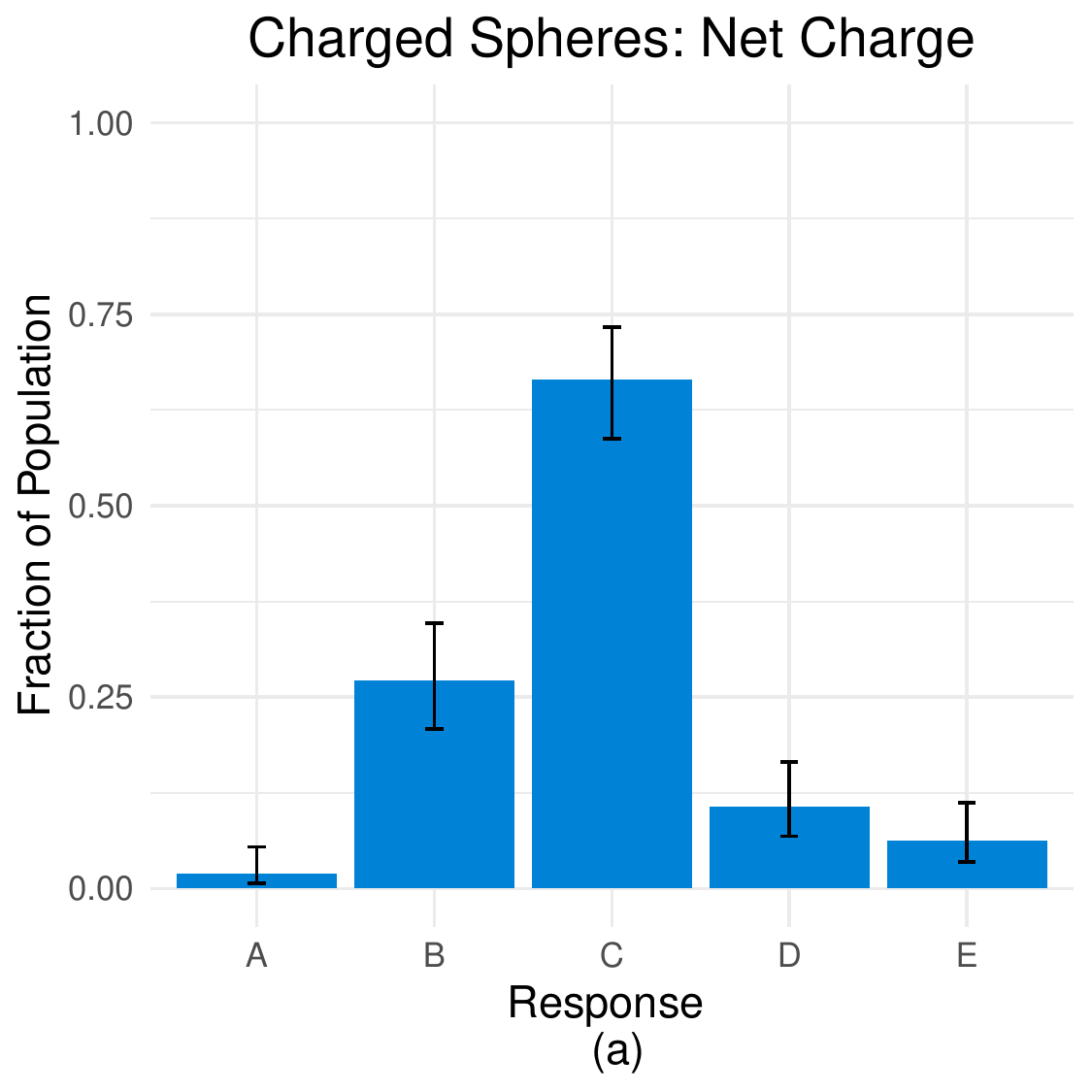}
    \hspace{0.05\textwidth}
    \includegraphics[width = 0.45\textwidth]{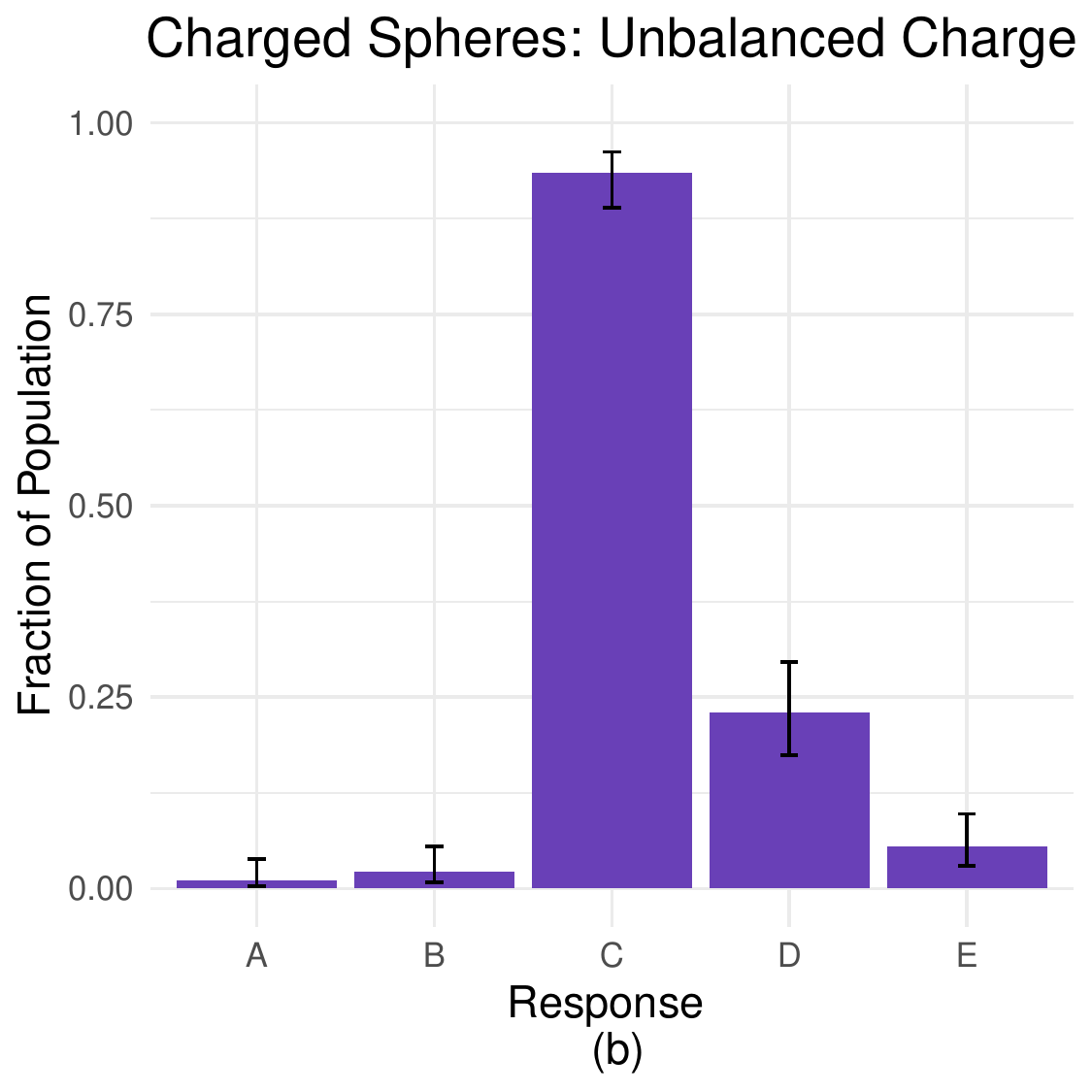}
    \caption{Response frequencies for two versions of the \textit{Charged Spheres} question shown in Fig.\ \ref{fig:ChargedSpheres}. Note: values do not add to unity because students may choose as many or as few answers as they think are correct. Responses C and D are both correct for both versions. Error bars represent the 95\% confidence intervals calculated using the Wilson method for binomial distributions \cite{binom, r, ggplot}.}
    \label{fig:sphereRes}
\end{figure*}

Results from the Charged Spheres question are summarized in Figure \ref{fig:sphereRes}. On version 1, 67\% of students (105 of $N=158$) included answer choice C, that the net charge on sphere A is neither greater than nor less than the net charge on sphere B. On version 2, however, 93\% of students (171 of $N=183$) included choice C, that the two spheres have equal amounts of unbalanced charge. (A binomial test indicates that this difference is statistically significant, with $p <0.001$.) The fraction of students selecting incorrect choice B (which is consistent with number-line reasoning) was larger on version 1 than on version 2 (27\% compared with 2\%). The fraction of students who selected choice D was twice as great on version 2 as compared to version 1 (23\% vs. 10\%), even though the language of this answer choice (as well as the related but mutually exclusive choice E) is identical in the two versions. (Again, a binomial test indicates this difference is statistically significant, with $p <0.001$.) Selecting choice D requires students to recognize that electrons (not protons) are moving from one sphere to the other.  We also find that on both versions of the question, the fraction of students selecting choice C was greater than the fraction selecting choice D.  (Recall that both choice C and choice D are correct.)

These results suggest that the wording differences in the two versions can indeed affect student performance. More descriptive wording, designed to prompt magnitude-based reasoning about charge as a signed quantity, was associated with a larger fraction of students recognizing values of charge with opposite sign but the same absolute value as corresponding to equal amounts.  In contrast, the less descriptive language, involving only the term ``net charge,'' seemed to be associated with interpretation of values of charge as positions on a number line (reasoning we consider inappropriate for electric charge).

\section{Instructional implications}
\label{sec:InstrImplications}
Reasoning about the concept of electric charge, and of ``net charge'' in particular, presents a greater learning challenge than students and instructors might initially recognize, in part due to subtleties in the use of positive and negative signs to characterize complementary charges. We offer three suggestions for promoting student learning of charge, anticipating that expert instructors will also devise their own approaches. 

\begin{enumerate}
\item \textbf{Include explicit language involving ``unbalanced charge''} Many instructors and textbooks use the phrase ``magnitude of the net charge'' when asking students to consider the amount of charge in surplus.  We view this as a missed opportunity to emphasize the subtle and usually implicit interpretation of sign as the signifier of type in the context of electric charge.  Our findings suggest that students may not spontaneously recognize that $-5$ and $+5$ indicate states with equal amounts of unbalanced charge.  We suggest instructors make use of the phrase ``amount of unbalanced charge'' as a way to help students interpret net charge in the intended manner. When the term ``net charge'' is used---without any other elaboration---some students may interpret the sign as indicating position on a number line.  The phrase ``amount of unbalanced charge,'' in contrast, seems to be a stronger cue, at least for some students, to compare inherently positive quantities (i.e., amounts).

\item \textbf{Discourage assumption of ``positivity''}: We caution against an assumption of positivity when discussing charge. We suggest that instructors specify ``$+5~\mu\textrm{C}$'' (rather than simply ``$5~\mu\textrm{C}$'') when discussing a positive net charge. The sign of any quantity carries meaning. For electric charge, the sign specifies the type, which in turn determines how the object will interact with other charged objects. Priming students to expect that real-world quantities have associated signs that carry meaning, and that an unsigned quantity is different than a positively-signed quantity, can help establish a physics habit-of-mind of actively seeking meaning in the sign. We also believe clearly labeling positive and negative charge with sign will aid in student recognition that variable or unknown amounts of charge (``$Q$'') could be either positive or negative \cite{brahmia2020}.

\item \textbf{Explicit instruction on zero-sum pairs}: We recognize that many students may not have considered ways of conceptualizing integers outside of positions on a number line. Because we believe the number-line strategy can be an obstacle when it is applied to electric charge, we suggest explicit instruction on the quantification of opposites as zero-sum pairs (i.e., pairs of numbers which sum to zero). There are a number of curricula that present strategies for understanding integers with a focus on zero-sum pairs \cite{thompson1998}. 
Explicit instruction in positive and negative numbers as complements rather than positions on either side of zero on a number line may help students understand better the quantification of electric charge.

\end{enumerate}

\bibliography{tpt_neg.bib}

\begin{thebibliography}{10}

\bibitem{brahmia2019b}
Suzanne~White Brahmia.
\newblock Quantification and its importance to modeling in introductory
  physics.
\newblock {\em European Journal of Physics}, 40(4):044001, 2019.

\bibitem{arons1990}
Arnold~B Arons and Thomas~D Miner.
\newblock A guide to introductory physics teaching, 1990.

\bibitem{vlassis2004}
Jo{\"e}lle Vlassis.
\newblock Making sense of the minus sign or becoming flexible in `negativity'.
\newblock {\em Learning and instruction}, 14(5):469--484, 2004.

\bibitem{brahmia2020}
Suzanne White~Brahmia, Alexis Olsho, Trevor~I. Smith, and Andrew Boudreaux.
\newblock Framework for the natures of negativity in introductory physics.
\newblock {\em Phys. Rev. Phys. Educ. Res.}, 16:010120, Apr 2020.

\bibitem{bishop2014}
Jessica~Pierson Bishop, Lisa~L Lamb, Randolph~A Philipp, Ian Whitacre, Bonnie~P
  Schappelle, and Melinda~L Lewis.
\newblock Obstacles and affordances for integer reasoning: An analysis of
  children's thinking and the history of mathematics.
\newblock {\em Journal for Research in Mathematics Education}, 45(1):19--61,
  2014.

\bibitem{brahmia2017a}
Suzanne Brahmia and Andrew Boudreaux.
\newblock Signed quantities: Mathematics based majors struggle to make meaning.
\newblock In {\em Proceedings of the 20th Annual Conference on RUME}, 2017.

\bibitem{binom}
Sundar Dorai-Raj.
\newblock {\em binom: Binomial Confidence Intervals For Several
  Parameterizations}, 2014.
\newblock R package version 1.1-1.

\bibitem{r}
{R Core Team}.
\newblock {\em R: A Language and Environment for Statistical Computing}.
\newblock R Foundation for Statistical Computing, Vienna, Austria, 2018.

\bibitem{ggplot}
Hadley Wickham.
\newblock {\em ggplot2: Elegant Graphics for Data Analysis}.
\newblock Springer-Verlag New York, 2016.

\bibitem{thompson1998}
Frances~McBroom Thompson et~al.
\newblock {\em Hands-on algebra!: Ready-to-use games \& activities for grades
  7-12}, volume~4.
\newblock Jossey-Bass, 1998.

\end{thebibliography}
\bibliographystyle{unsrt} 
\printendnotes

\end{document}